\documentclass[twocolumn]{article}
\pdfoutput=1
\usepackage[a4paper,vmargin=2cm,hmargin=1.5cm]{geometry}
\usepackage{sectsty}
\usepackage{graphicx}
\usepackage{amsmath}
\usepackage{url}
\usepackage{booktabs} 
\usepackage{xspace}
\usepackage{gensymb} 
\usepackage{ragged2e}
\usepackage{mathtools}
\usepackage{xspace}
\usepackage[font=footnotesize,justification=justified,format=plain]{caption}

\usepackage[T1]{fontenc}
\usepackage{newtxmath,newtxtext}

\sectionfont{\fontsize{12}{15}\selectfont}

\newcommand{\ssimA}[5]{ %
	\setlength{\fboxsep}{0pt}%
	\setlength{\fboxrule}{0.2mm}%
	{\footnotesize%
		$\mathrm{MSSIM} \left(\begin{tabular}{c} \ \\ \ \\ \ \\ \ \end{tabular} \right. $ %
		\hspace*{-6.0mm} %
		\raisebox{-1.6mm}{\begin{tabular}{c} %
				\fbox{\includegraphics[width=0.065\textwidth]{#1}} \\%
				#2 %
		\end{tabular}}\hspace*{-0.9mm}%
		,
		\hspace{-1.3mm}
		\raisebox{-1.6mm}{\begin{tabular}{c} %
				\fbox{\includegraphics[width=0.065\textwidth]{#3}}  \\%
				#4 %
		\end{tabular}}\hspace{-4.8mm}%
		$\left.\begin{tabular}{c} \ \\ \ \\ \ \\ \ \end{tabular} \right) =\, $#5%
}}

\newcommand{\ssimB}[6]{ %
	\setlength{\fboxsep}{0pt}%
	\setlength{\fboxrule}{0.2mm}%
	{\footnotesize%
		$\mathrm{SSIM} \left(\begin{tabular}{c} \ \\ \ \\ \ \\ \ \end{tabular} \right. $ %
		\hspace*{-6.0mm} %
		\raisebox{-1.8mm}{\begin{tabular}{c} %
				\fbox{\includegraphics[width=0.065\textwidth]{#1}} \\%
				#2 %
		\end{tabular}}\hspace*{-0.9mm}%
		, 
		\hspace{-1.3mm}
		\raisebox{-1.8mm}{\begin{tabular}{c} %
				\fbox{\includegraphics[width=0.065\textwidth]{#3}}  \\%
				#4 %
		\end{tabular}}\hspace{-4.8mm}%
		$\left.\begin{tabular}{c} \ \\ \ \\ \ \\ \ \end{tabular} \right) =\, $%
		\hspace{-1.3mm}%
		\raisebox{-1.8mm}{\begin{tabular}{c} %
				\fbox{\includegraphics[width=0.065\textwidth]{#5}}  \\%
				#6 %
		\end{tabular}}%
}}

\newcommand{\componentWidth}{0.10\columnwidth}
\newcommand{\ccell}[1]{\begin{tabular}{@{}c@{}} #1 \end{tabular}}
\newcommand{\iccell}[1]{\ccell{\fbox{\includegraphics[width=\componentWidth]{#1}}} }

\newcommand{\ssimD}{
	\setlength{\fboxsep}{0pt}
	\setlength{\fboxrule}{0.2mm}
	\setlength\tabcolsep{1pt}
	{\scriptsize%
		\begin{tabular}{cccccccccccccc}
			\ccell{SSIM} & $\left(\hspace{1mm}\iccell{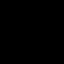}\right.$ & \ccell{,} & $\left.\iccell{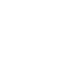}\hspace{1mm}\right)$ & = & \iccell{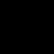} & $\!\!\cdot\!$ & \iccell{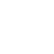} & $\!\!\cdot\!$ & \iccell{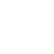} & = & \iccell{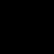} \\
			& \hspace{2.5mm}0/255 &  & \hspace{-1mm}255/255 &	& $l=0.0001$ & & $c=1$ & & $s=1$ & & $0.0001$ \\
			\multicolumn{12}{c}{\rule{\columnwidth}{0.2mm}} \\
			\ccell{SSIM} & $\left(\hspace{1mm}\iccell{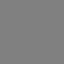} \right.$ & \ccell{,} & $\left.\iccell{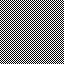}\hspace{1mm}\right)$ & = & \iccell{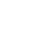} & $\!\!\cdot\!$ & \iccell{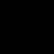} & $\!\!\cdot\!$ & \iccell{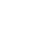} & = & \iccell{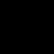} \\
			& \hspace{2.5mm}128/255 &  & \hspace{-1mm}b|w &	& $l=1$ & & $c=0.0036$ & & $s=1$ & & $0.0036$ \\
			& \hspace{2.5mm}\iccell{images/gray/gray128.png} & & \hspace{-1mm}\iccell{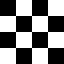} \\
			& \hspace{2.5mm}$16\times$ & & \hspace{-1mm}$16\times$ \\
			\multicolumn{12}{c}{\rule{\columnwidth}{0.2mm}} \\
			\ccell{SSIM} & $\left(\hspace{1mm}\iccell{images/checker-bw.png} \right.$ & \ccell{,} & $\left.\iccell{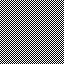}\hspace{1mm}\right)$ & = & \iccell{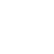} & $\!\!\cdot\!$ & \iccell{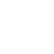} & $\!\!\cdot\!$ & \iccell{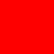} & = & \iccell{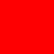} \\
			& \hspace{2.5mm}b|w &  & \hspace{-1mm}w|b &	& $l=1$ & & $c=1$ & & $s=-0.9964$ & & $-0.9964$ \\
			& \hspace{2.5mm}\iccell{images/checker-bw_16x_crop.png} & & \hspace{-1mm}\iccell{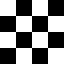} \\
			& \hspace{2.5mm}$16\times$ & & \hspace{-1mm}$16\times$ \\
		\end{tabular}
	}
}

\newcommand{\img}[1]{{\textbf{#1}}}
\newcommand{\muA}{\mu_\img{A}}
\newcommand{\muB}{\mu_\img{B}}
\newcommand{\sigmaA}{\sigma_\img{A}}
\newcommand{\sigmaB}{\sigma_\img{B}}
\newcommand{\sigmaAB}{\sigma_{\img{A}\img{B}}}

\makeatletter
\newcommand{\customlabel}[2]{%
	\protected@write \@auxout {}{\string \newlabel {#1}{{#2}{}}}}
\makeatother

\newcommand{\FLIP}{\protect\reflectbox{F}LIP\xspace}

\title{Understanding SSIM}
\author{Jim Nilsson \\ NVIDIA \and Tomas Akenine-M\"oller \\ NVIDIA}
\date{%
		\setcounter{figure}{1}
		\begin{center}
			\setlength{\fboxsep}{0pt}%
			\setlength{\fboxrule}{0.2mm}%
			\setlength{\tabcolsep}{0pt}
			\renewcommand{\arraystretch}{0.2}
			\newcommand{\teaserImageHeight}{0.13\textwidth}
			\newcommand{\smallTeaserImageHeight}{0.0418\textwidth}
			\footnotesize
			\begin{tabular}{ccccccccccccccccccccc}
				& \multicolumn{3}{c}{\textit{a}} & \multicolumn{3}{c}{\textit{b}} & \multicolumn{3}{c}{\textit{c}}
				& \multicolumn{3}{c}{\textit{d}} & \multicolumn{3}{c}{\textit{e}} & \multicolumn{3}{c}{\textit{f}} &
				\\
				\rotatebox{90}{\hspace*{3.2mm}reference}
				& \multicolumn{3}{c}{\includegraphics[height=\teaserImageHeight]{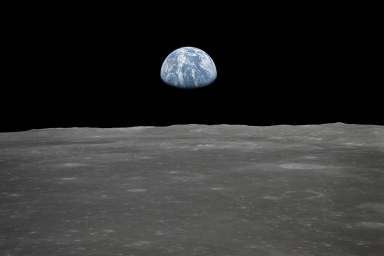}} 
				& \multicolumn{3}{c}{\includegraphics[height=\teaserImageHeight]{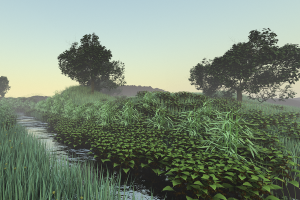}}
				& \multicolumn{3}{c}{\includegraphics[height=\teaserImageHeight]{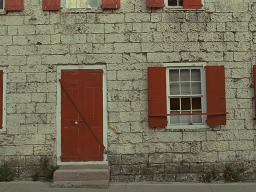}}
				& \multicolumn{3}{c}{\includegraphics[height=\teaserImageHeight]{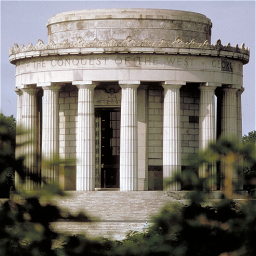}}
				& \multicolumn{3}{c}{\includegraphics[height=\teaserImageHeight]{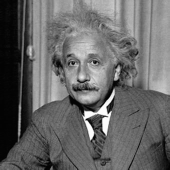}}
				& \multicolumn{3}{c}{\includegraphics[height=\teaserImageHeight]{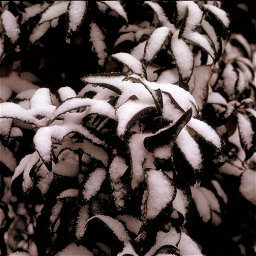}}
				&
				\\ 
				\rotatebox{90}{\hspace*{8mm}test}
				& \multicolumn{3}{c}{\includegraphics[height=\teaserImageHeight]{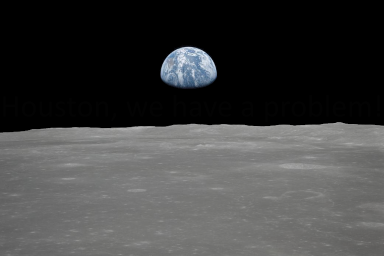}}
				& \multicolumn{3}{c}{\includegraphics[height=\teaserImageHeight]{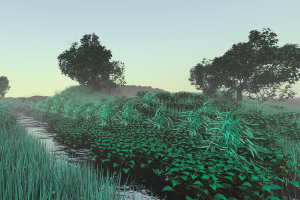}}
				& \multicolumn{3}{c}{\includegraphics[height=\teaserImageHeight]{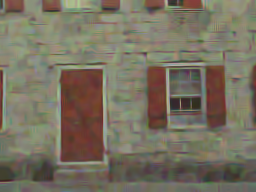}}
				& \multicolumn{3}{c}{\includegraphics[height=\teaserImageHeight]{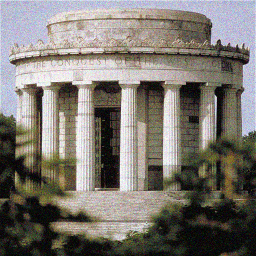}}
				& \multicolumn{3}{c}{\includegraphics[height=\teaserImageHeight]{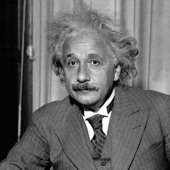}}
				& \multicolumn{3}{c}{\includegraphics[height=\teaserImageHeight]{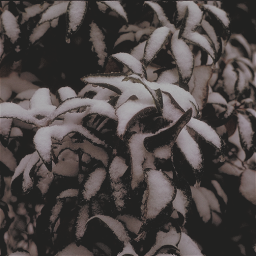}}
				&
				\\ 
				\rotatebox{90}{\hspace*{5mm}SSIM}
				& \multicolumn{3}{c}{\includegraphics[height=\teaserImageHeight]{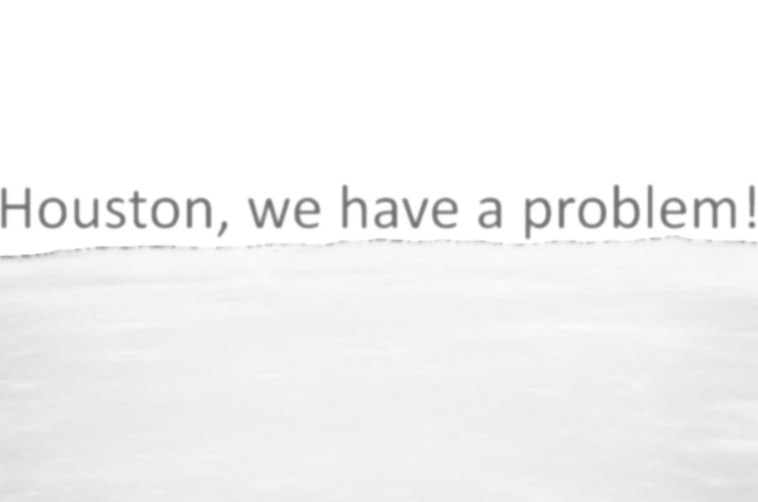}}
				& \multicolumn{3}{c}{\includegraphics[height=\teaserImageHeight]{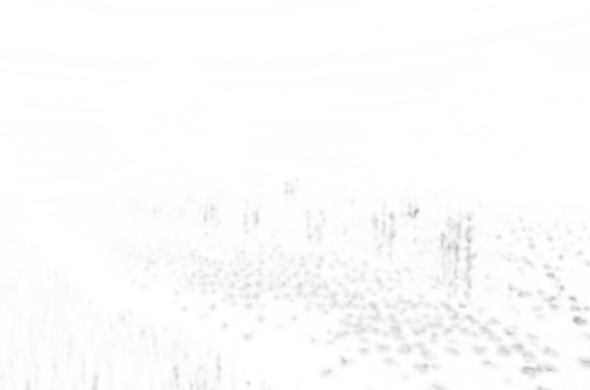}}
				& \multicolumn{3}{c}{\includegraphics[height=\teaserImageHeight]{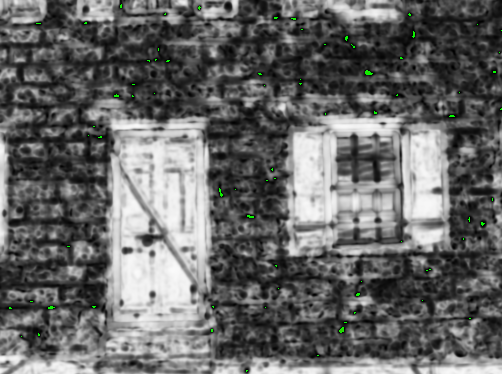}}
				& \multicolumn{3}{c}{\includegraphics[height=\teaserImageHeight]{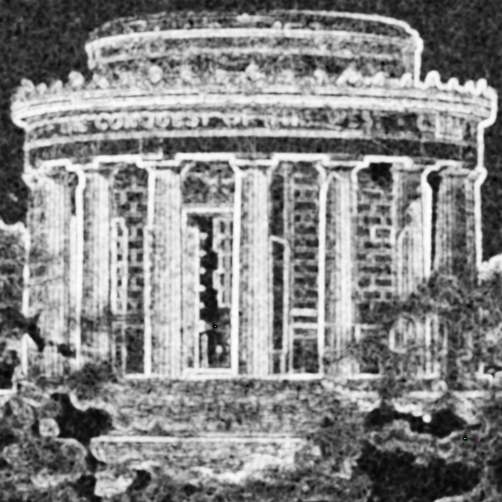}}
				& \multicolumn{3}{c}{\includegraphics[height=\teaserImageHeight]{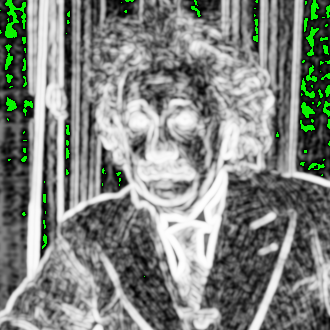}}
				& \multicolumn{3}{c}{\includegraphics[height=\teaserImageHeight]{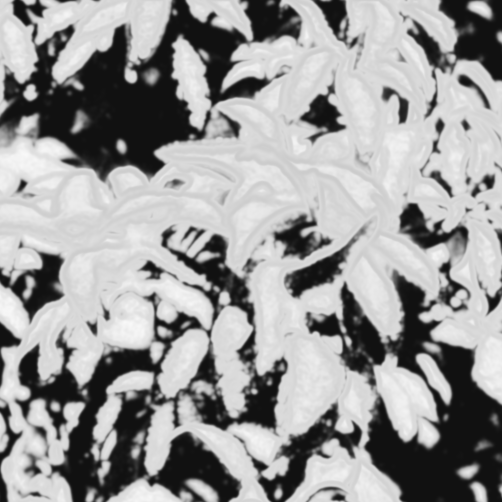}}
				& \fbox{\includegraphics[height=\teaserImageHeight]{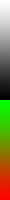}}
				\\ 
				& \fbox{\includegraphics[height=\smallTeaserImageHeight]{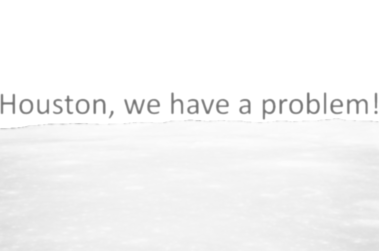}}
				& \fbox{\includegraphics[height=\smallTeaserImageHeight]{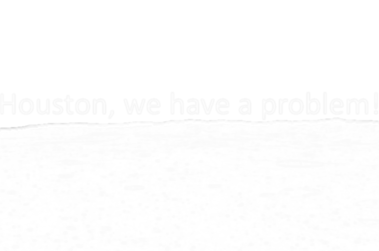}}
				& \fbox{\includegraphics[height=\smallTeaserImageHeight]{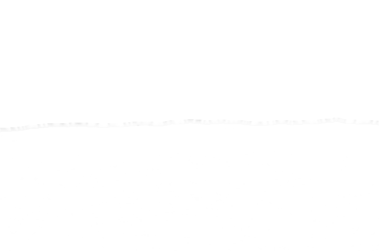}}
				& \fbox{\includegraphics[height=\smallTeaserImageHeight]{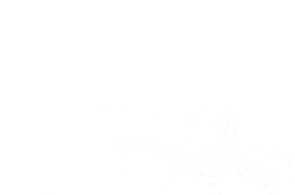}}
				& \fbox{\includegraphics[height=\smallTeaserImageHeight]{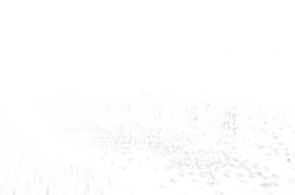}}
				& \fbox{\includegraphics[height=\smallTeaserImageHeight]{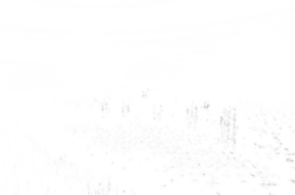}}
				& \fbox{\includegraphics[height=\smallTeaserImageHeight]{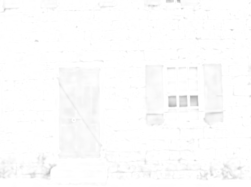}}
				& \fbox{\includegraphics[height=\smallTeaserImageHeight]{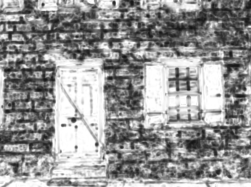}}
				& \fbox{\includegraphics[height=\smallTeaserImageHeight]{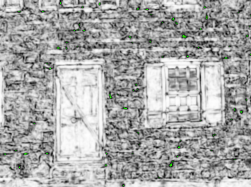}}
				& \fbox{\includegraphics[height=\smallTeaserImageHeight]{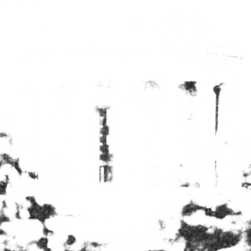}}
				& \fbox{\includegraphics[height=\smallTeaserImageHeight]{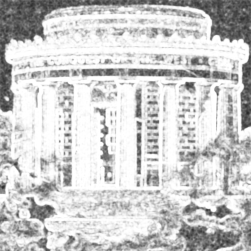}}
				& \fbox{\includegraphics[height=\smallTeaserImageHeight]{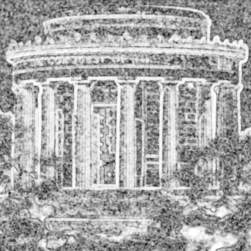}}
				& \fbox{\includegraphics[height=\smallTeaserImageHeight]{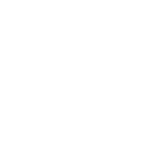}}
				& \fbox{\includegraphics[height=\smallTeaserImageHeight]{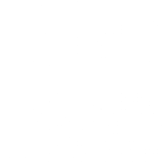}}
				& \fbox{\includegraphics[height=\smallTeaserImageHeight]{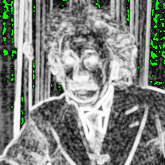}}
				& \fbox{\includegraphics[height=\smallTeaserImageHeight]{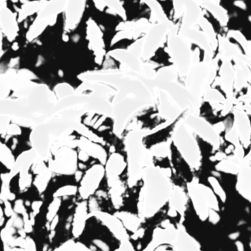}}
				& \fbox{\includegraphics[height=\smallTeaserImageHeight]{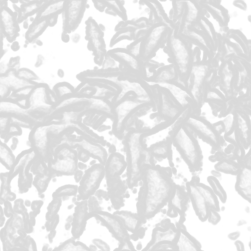}}
				& \fbox{\includegraphics[height=\smallTeaserImageHeight]{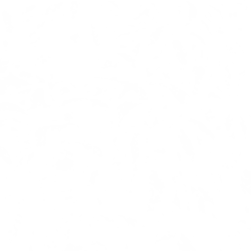}}
				&
				\\
				&
				$l$ & $c$ & $s$ &
				$l$ & $c$ & $s$ &
				$l$ & $c$ & $s$ &
				$l$ & $c$ & $s$ &
				$l$ & $c$ & $s$ &
				$l$ & $c$ & $s$ &
			\end{tabular}
		\end{center}%
		\vspace*{-3mm}
		\begin{justify}
			\footnotesize
			Figure \thefigure: Images should ideally be viewed
			on a display at 100\% scale, so
			we urge the reader to look at the images in our supplemental material\footnote{https://research.nvidia.com/publication/2020-07\_Understanding-SSIM}.
			The SSIM images (3rd row) are generated from the references (1st row) and the test images (2nd row),
			and $\mathrm{SSIM}$ is visualized using 
			the heatmap to the lower right, where white is $\mathrm{SSIM}=1$ (identical images), black is $\mathrm{SSIM}=0$,
			and SSIM values in $(-1,0]$ map to $(\mathrm{red},\mathrm{green}]$.
			The 4th row contains the $l$, $c$, and $s$ components of SSIM\@.
			Comments:
			\textit{a)} the dark text is detected,
			\textit{b)} chrominance differences are hardly noticed,
			\textit{c)} large areas on the red door and window shutter are not detected,
			\textit{d)} edges in images often generate an SSIM value close to one (white),
			even though the error is large all over this test image,
			\textit{e)} the Einstein images are similar unless highly zoomed-in, but SSIM reports large errors, and
			\textit{f)} the final pair where the SSIM image is leaning toward being almost inverted
			compared to what one may expect.
			Note that all reference and test images, but not SSIM images, are reduced in size for space considerations.
		\end{justify}
		\vspace{\fill}
		\customlabel{fig_teaser}{\thefigure}
}

\begin{document}
	
	\maketitle

	\section*{Abstract}
	The use of the structural similarity index (SSIM) is widespread.
	For almost two decades, it has played a major role in image quality
	assessment in many different research disciplines.
	Clearly, its merits are indisputable in the research community.
	However, little deep scrutiny of this index has been performed.
	Contrary to popular belief, there are some interesting properties of SSIM that merit such scrutiny.
	In this paper, we analyze the mathematical factors of SSIM and show
	that it can generate results, in both synthetic and realistic use cases,
	that are unexpected, sometimes undefined, and nonintuitive.		
	As a consequence, assessing image quality based on SSIM can lead to incorrect
	conclusions and using SSIM as a loss function for deep learning can guide neural network
	training in the wrong direction.
	
	\section{Introduction}
	\label{sec_intro}
	
	The original SSIM paper~\cite{Wang2004a} has over $20,\!000$ citations on Google Scholar.
	Thousands of research papers have used it as a quality index when comparing images,
	and we are indeed authors of a few of those.
	In this paper, we provide a review and a deep
	inspection of SSIM, and we show that SSIM can
	deliver unexpected or invalid results in both simple use cases and for real image pairs. See Figure~\ref{fig_teaser}.
	We start with an overview of SSIM\@.
	
	The input color space of SSIM is never defined.
	As the reference Matlab script performs no color space transformations on inputs,
	our assumption throughout this paper is that all images are encoded 
	in sRGB color space, i.e., approximately gamma encoded with an exponent $\approx 2.4$.
	Note that this means that an image that is loaded by the SSIM script is assumed to be viewed directly on screen as is.
	For two images $\img{A}$ and $\img{B}$,
	the original formula~\cite{Wang2004a} for per-pixel SSIM is given by
	\begin{equation}
	\begin{aligned}
	\mathrm{SSIM}(x,y) &= \Big(l(x,y)\Big)^\alpha  
	\Big(c(x,y)\Big)^\beta  \Big(s(x,y)\Big)^\gamma,
	\end{aligned}
	\label{eq_SSIM}
	\end{equation}
	where $\img{A}$ and $\img{B}$ are inputs to all functions, but omitted for clarity.
	To compute mean, variance, and covariance in a patch around a pixel, they 
	use Gaussian-weighted versions of these formulae with a filter kernel of $11\times 11$ pixels and $\sigma=1.5$.
	The \textit{luminance} component, $l$, is then	
	\begin{equation}
	l(x,y) = \frac{ 2\muA \muB + C_1}{\muA^2 +\muB^2 + C_1} ,
	\label{eq_SSIM_luminance}
	\end{equation}
	where the mean values $\muA$ and $\muB$ are functions of $x,y$ as well, e.g.,  $\muA(x,y)$,
	but we skip the $(x,y)$ in favor of shorter notation, and likewise for the variances
	$\sigmaA^2$ and $\sigmaB^2$, and for the
	covariance $\sigma_{\img{A}\img{B}}$.
	The \textit{contrast} component, $c$, is
	\begin{equation}
	c(x,y) = \frac{ 2 \sigmaA\sigmaB + C_2 }{\sigmaA^2 + \sigmaB^2 + C_2}.
	\label{eq_SSIM_contrast}
	\end{equation}
	Finally, the \textit{structure} component, $s$, is
	\begin{equation}
	s(x,y) = \frac{\sigma_{\img{A}\img{B}} + C_3} {\sigmaA\sigmaB + C_3}.
	\label{eq_SSIM_structure}
	\end{equation}
	Wang et al.~propose that $C_1=(K_1 L)^2$, $C_2=(K_2 L)^2$, and $C_3=C_2/2$, where $L=255$ for 8-bit component images.
	Furthermore, they chose $K_1=0.01$ and $K_2=0.03$. If the range for an image is $[0,1]$,
	we set $L=1$ in order to get the same result, i.e., $C_1 = K_1^2$ and $C_2=K_2^2$. Finally,
	the mean SSIM (MSSIM) value, which is pooled over the entire image, is
	\begin{equation}
	\mathrm{MSSIM}(\img{A},\img{B}) = \frac{1}{wh} \sum_x \sum_y \mathrm{SSIM}(x,y),
	\end{equation}
	where $w$ and $h$ are the width and height of the image.
	As can be seen, the term $\sigmaA\sigmaB$ is in the numerator in Equation~\ref{eq_SSIM_contrast}
	and in the denominator in Equation~\ref{eq_SSIM_structure}.
	To create a simplified expression, 
	Wang et al.~therefore proposed to use $\alpha=\beta=\gamma=1$ and $C_3=C_2/2$,
	which results in
	\begin{equation}
	\begin{aligned}
	\mathrm{SSIM}(x,y) &= 
	\frac{ ( 2\muA \muB + C_1) (2\sigmaAB + C_2) }
	{ (\muA^2 +\muB^2 + C_1) (\sigmaA^2 + \sigmaB^2 + C_2) }.
	\end{aligned}
	\label{eq_SSIM_simplified}
	\end{equation}
	Next, we review some related work.
	
\section{The History of SSIM}
\label{sec_prev_work}

The universal quality index (UQI), which was introduced by Wang and Bovik~\cite{Wang2002},
is essentially SSIM as presented above, but without any of the
constants $C_i$. These constants were added later to avoid
division by zero~\cite{Wang2004a}.
Multi-scale SSIM (MS-SSIM)\footnote{Note the
	difference between MSSIM, which is the average SSIM value over an image pair, and 
	MS-SSIM, which is the multi-scale variant of SSIM\@.}
was introduced as a means for including image details at different scales~\cite{Wang2003}. MS-SSIM adds more
components in the expression, where both the contrast and structure expressions
are evaluated at five low-pass filtered and downsampled versions of the original images.
However, the components have the same form as in SSIM. 

Sampat et al.~\cite{Sampat2009} introduce complex wavelet structural similarity
(CW-SSIM), where the expression in Equation~\ref{eq_SSIM_simplified}
is used, but where the components are replaced by complex wavelet coefficients.
CW-SSIM is more tolerant to small translations and rotations, which may be
a desired effect in some contexts. However, for rendered images, which often contain geometrical edges, it is most likely
not a desired feature, since, for instance, a game designer usually wants the geometry to be precisely where he/she intends.
3D-SSIM~\cite{Zeng2012} is an extension of SSIM for video, where the formulae 
are evaluated for three-dimensional blocks of pixel values and multiplied
with information content weights and local distortion weights.
SSIM is still being adapted for new uses, e.g.,
spherical SSIM~\cite{Chen2018}, where SSIM was adapted to handle a spherical projection,
and for medical images~\cite{Liu2018}. 

While mean square error (MSE) has been criticized~\cite{Wang2009} for not delivering a truthful value compared to image error,
Dosselman and Yang~\cite{Dosselmann2011} and Hor\'e and Ziou~\cite{Hore2010} have, at the same time,
shown that there is a close relationship between MSE and SSIM\@.
Whittle et al.~\cite{Whittle2017} evaluate image metrics for Monte Carlo
rendered images with different levels of noise. Their conclusion
is that MS-SSIM performs well for this task.
{\v C}ad\'{\i}k et al.~\cite{Cadik2012} perform an extensive evaluation
of image indices and metrics together with a user study, and find
contradictory results for several of the algorithms, including SSIM,
for various image distortions.  Recently, SSIM has found uses as a loss function
for deep learning~\cite{Zhao2017}, and is also included in tool kits
such as Tensorflow.

Neither the UQI nor SSIM make any claim to be a \emph{metric} in the mathematical sense,
for which the triangle inequality, i.e., $d(x,z) \le d(x,y) + d(y,z)$, must hold.
It has, however, been shown that $\sqrt{1 - l(x,y)}$ and $\sqrt{1 - c(x,y)s(x,y)}$ do fulfill the triangle inequality,
and thus are metrics~\cite{Brunet2012}.
Interestingly, 
the derivation to transform a modified version of SSIM into a metric, also revealed subtle---while
important---properties of the index.
For instance, for $l(x, y)$, we have:
\begin{equation}
\sqrt{1 - l(x,y)} = \frac{|\muA - \muB|}{\sqrt{\muA^2 +\muB^2 + C_1}},
\end{equation}
which can be seen as a normalized version of the root mean square error (RMSE)\@.

This is notable, since MSE is not a perception-based metric~\cite{Girod1993,Wang2009} and the finding above
has the implication that there could exist a direct relationship between SSIM and MSE, which has indeed been
independently discovered by Dosselman and Yang~\cite{Dosselmann2008,Dosselmann2011}
and Hor\'e and Ziou~\cite{Hore2010}.
Specifically, Dosselman and Yang show that there is a direct mathematical transform between $\text{SSIM}^\ast$
and $\text{MSE}^\ast$. 
$\text{SSIM}^\ast$ is SSIM with constants $C_i=0$,
which does not alter the validity of the analysis, while $\text{MSE}^\ast$ is a local MSE, using the same footprint as SSIM\@.
They further show this empirically by correlating MSE versus SSIM for a range
of images, using the coefficient of multiple determination, $R^2$, which is $0.0$ for no association between
the variables and values closer to $1.0$ indicate strong degree of correspondence~\cite{Dosselmann2008}. It was found that $R^2$ was between
$0.9322$ and $1.0$, which implies that $\text{SSIM}^\ast$ and $\text{MSE}^\ast$ perform similarly.

Furthermore, Hor\'{e} and Ziou~\cite{Hore2010} make a similar discovery, i.e., that there is a close relationship
between $\text{SSIM}^\ast$, peak-signal-to-noise-ratio (PSNR), and MSE\@.  They identify that $\text{MSE} = \sigmaA^2 +\sigmaB^2 - 2\sigmaAB + (\muA - \muB)^2$,
and can then derive PSNR as a function of $\text{SSIM}^\ast$.  For images of similar luminance, i.e., $\muA \approx \muB$,
and SSIM values in the $[0.2, 0.8]$ range, this function is approximately linear, which indicates that for any other distortion
than a luminance shift, $\text{SSIM}^\ast$ is qualitatively equivalent to PSNR\@.

These findings question the validity of claims that SSIM is
a perception-based index, since MSE is not a perception-based metric~\cite{Girod1993,Wang2009}. 
The small discrepancies in correlation between MSE and SSIM were shown to stem partly from the fact that
SSIM is derived for a spatial subregion of the whole images, and an effect of the $C_i$ constants~\cite{Dosselmann2008,Dosselmann2011}.

	\section{Mathematical Properties}
	\label{sec_mathematics}
	
	This section will analyze the components of SSIM from a mathematical standpoint.
	The behavior of the quality index itself will be scrutinized in the next section.
	
	We use the notation $x/y$ to denote a pixel
	of a particular grayscale value.  For example, $128/255$
	corresponds to a gray value of 50.2\%.  The shorthand form $x|y$ designates a
	pixel-sized checkerboard pattern with colors $x/255$ and $y/255$.  Letters \emph{b}
	and \emph{w} are shorthand for black ($0/255$) and white ($255/255$), respectively.
	
	
	\subsection{Minimum Values of the SSIM Factors}
	\label{sec_min_values}
	To understand the workings of the components of SSIM, as we will see later in this section,
	and since it, to our knowledge, has not been done before,
	we explain how to minimize $l$, $c$, and $s$,
	one at a time. 
	For $l(x, y)$, shown in Equation~\ref{eq_SSIM_luminance}, we can
	differentiate and solve for zero, with the assumptions that $\mu_\img{A}, \mu_\img{B}\in [0,L]$.
	This gives us a minimum when $\mu_\img{A}=0$ and $\mu_\img{B}=L$ (or vice versa). 
	Hence, we have  $l_\mathrm{min} = C_1 / (L^2 + C_1) = K_1^2 / (K_1^2 + 1) = 0.0001$, for $K_1=0.01$.
	
	The $c$ and $s$ components are slightly more involved and first we note
	that maximum variance and covariance for variables in $[0,L]$ is $(L/2)^2$.
	For the $c$ component (Equation~\ref{eq_SSIM_contrast}), we can differentiate in the same manner as for $l$,
	and find that the minimum occurs when $\sigmaA=0$ and $\sigmaB=(L/2)^2$ (or vice versa), which gives
	$c_\mathrm{min} = C_2 / (L^2 / 4 + C_2) = K_2^2 / (K_2^2 + 0.25) = 0.0036$, for $K_2=0.03$.
	Brunet~\cite{Brunet2012b} has ascertained that $s\in[-1,1]$, but one can do slightly better than that, since
	$s$ (Equation~\ref{eq_SSIM_structure}) is minimized for minimum covariance $\sigmaAB$ and maximum variances ($\sigmaA^2$, $\sigmaB^2$).
	This gives the minimum at $s_\mathrm{min}=(-L^2 / 4 + C_3) / (L^2 / 4 + C_3) = (2K_2^2 - 1)/(2K_2^2 +1) = -0.9964$.
	
	
	\begin{figure}[tb]
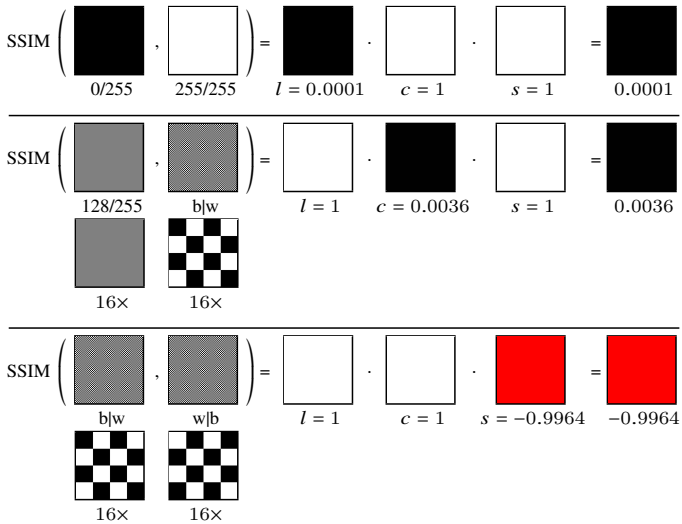

		\ssimD
		\caption{Minimum values of the SSIM components: $l_\mathrm{min} = 0.0001$ (top), $c_\mathrm{min} = 0.0036$ (middle), and $s_\mathrm{min} = -0.9964$ (bottom). The images at the bottom in the middle and bottom examples have been zoomed by a factor $16\times$.}
		\label{fig_LCS_minima}
	\end{figure}
	
	To give an example that the above minima can occur,
	we show examples of when SSIM becomes $l_\mathrm{min}$, $c_\mathrm{min}$, and $s_\mathrm{min}$
	in Figure~\ref{fig_LCS_minima}.
	The first row uses a black and a white image, which results
	in $l=l_\mathrm{min}$ and 
	$\sigmaA=$ $\sigmaB=$ $\sigmaAB=0$, implying that $c = s = 1$, which gives SSIM=$l_\mathrm{min}$.
	The second row uses
	a 128/255 image and a b$|$w checkerboard, which thus has $\muA=\muB$, 
	while at the same time making $\sigmaA=0$ and $\sigmaB=(L/2)^2$. As a result,
	SSIM=$c_\mathrm{min}$.
	Looking at the second row of Figure~\ref{fig_LCS_minima} with a low dot pitch,
	it is hard for the human visual system (HVS) to discern differences between the
	images, while SSIM values are close to zero, which indicates low quality contrary to the actual experience.
	The last row in Figure~\ref{fig_LCS_minima} achieves SSIM=$s_\mathrm{min}$
	by using two inverted checkerboard images.
	Note that all three minima are independent of $L$, as expected, and it is clear that the range of SSIM is $(-1, 1]$.
	
	While most people use the simplified version (Equation~\ref{eq_SSIM_simplified}) of SSIM,
	which is also what the reference implementation~\cite{Wang2004a} uses,
	the constants $\alpha$, $\beta$, and $\gamma$ have been used to find a more optimized
	SSIM expression~\cite{Piorowski2016} for antialiasing detection in games,
	but also in MS-SSIM~\cite{Wang2003}, and for optimizing SSIM parameters using machine learning~\cite{Cadik2013}.
	Therefore, it is important to take a look at the components
	in the full SSIM expression (Equation~\ref{eq_SSIM}) as well and see what their ranges are.
	The $s$ component (Equation~\ref{eq_SSIM_structure}) deserves additional attention.
	If $C_3$ is zero, $s(x,y)$ is equivalent to the sample Pearson correlation coefficient, usually denoted $r_{xy}$.
	We have found no scientific support that this coefficient correlates with human perception of ``structure.''
	As mentioned in the introduction, SSIM has selected
	$C_3=C_2/2 = (K_2L)^2 /2 = (0.03\cdot 1)^2 /2 = 0.00045$, for pixel values in the range $[0,1]$.
	Since by definition we have $\sigmaA\geq 0$ and $\sigmaB\geq 0$, the denominator will always be positive.
	The covariance term $\sigmaAB$ can take on negative values,
	which means that $s(x,y)$ can be negative.
	Note that raising a negative number to a positive, non-integer number, $\gamma$,
	results in a complex number (with a real and an imaginary part), which in practice, e.g., in programming languages,
	makes the result \textit{undefined}. The \texttt{std::pow()} function gives \texttt{NaN} (not a number)
	when a negative number is raised to a number (even to one).
	MS-SSIM~\cite{Wang2003} uses non-integer $\gamma$-values, as do the work of {\v C}ad\'{\i}k et al.~\cite{Cadik2013}
	and Pi{\'o}rkowski \& Mantiuk~\cite{Piorowski2016}, so it seems that this is not well-known.
	
	Undefined results (or complex numbers), unless properly defined, should not be an outcome of an image quality index.
	Even if the range of SSIM is allowed to be complex, there are no descriptions on how to interpret such values. 
	To our knowledge, this problem has never been identified before
	and means that SSIM implementations can generate undefined
	results for some inputs and parameter settings.

	\subsection{Perceptual Properties}
	The purpose of deriving these minima is not only out of curiosity, but also hints at a deeper problem
	with the index itself and the claims of it being based on perception.
	Referring to the two bottom examples in Figure~\ref{fig_LCS_minima}, it could be argued that detecting the
	difference between the images leading to a minimum value for $c$ and $s$ is hard
	(please look at the images in the supplemental material).
	Depending on the monitor dot pitch and viewing distance, any difference
	between the images $128/255$ and $b|w$ can be indistinguishable to a human viewer.
	The same is true for the bottom row, images $b|w$ and $w|b$.
	This property will be further investigated in Section~\ref{sec_dissection}.

	As far as we can see, the only perception related claims made in the SSIM paper~\cite{Wang2004a}
	is that the $l$ component is qualitatively consistent with Weber's law and that the $c$ component
	is consistent with the contrast-masking feature of the HVS\@.
	This is somewhat contradictory, since
	in the original UQI paper, the authors explicitly state that ``the new index is mathematically
	defined and no human visual system model is explicitly employed.''
	Weber's law~\cite{Kandel2013} states that the ratio between the difference in stimulus against a \emph{background} signal,
	to achieve the same psychophysical sensation, is approximately constant ($\frac{\Delta S}{S}=k$).
	Contrast-masking is the destructive interference between (transient) stimuli closely coupled in space and time~\cite{Legge1981}.
	Both these psychophysical phenomena are well known.
	However, these effects are both defined only \emph{within the same frame of reference}, i.e., when introducing stimuli onto some sort of background.
	Consequently, in the context of image comparisons, these effects hold true for a local pixel value
	against its background \emph{within} the same image,	
	and not true for variations \emph{between} images.
	
	\begin{figure}[tb]
		\centering
		\includegraphics[width=0.9\columnwidth]{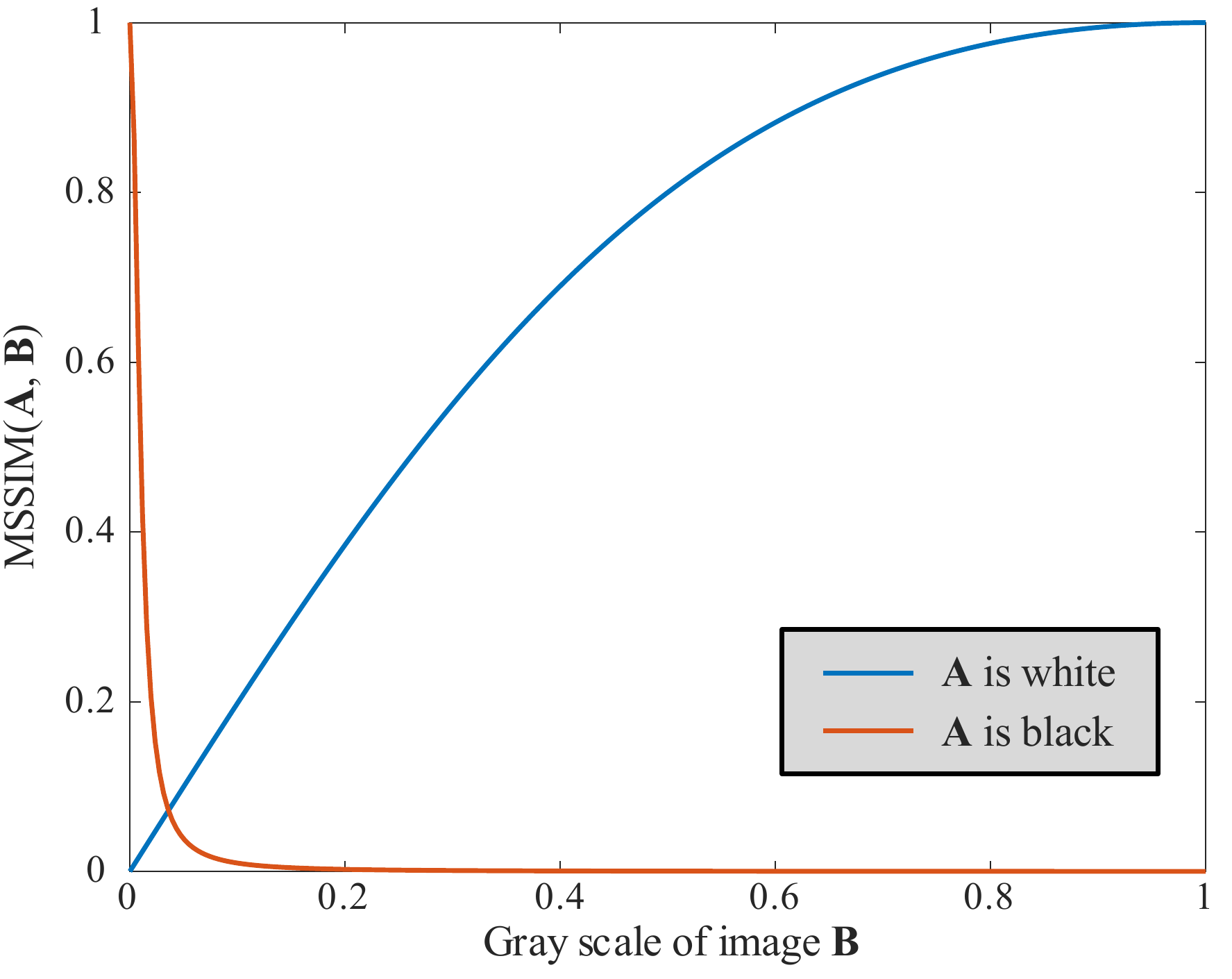}
		\caption{MSSIM as a function of an increasingly brighter, constant-valued image, $\img{B}$, 
			against a black image (orange curve) and a white image (blue curve).
		}
		\label{fig_black_against_gray}
	\end{figure}
	The claim that the $l$ factor is perceptually motivated is problematic.
	Evidence for this can easily be found and first, we point to Figure~\ref{fig_luminance_weirdnesses}
	to get a feel for this. To explain these results, we refer to
	Figure~\ref{fig_black_against_gray}, which shows a plot of SSIM
	as a function of a constant colored image from black to white, against
	both a black and a white image. Note that, as before, $c=s=1$.
	As can be seen, when the comparison is against a black image, $\img{A}$, and the image $\img{B}$ is close to black,
	a minuscule change in $\img{B}$ triggers a huge difference in $l$, and thus in SSIM\@.
	Furthermore, when $\img{A}$ is black and $\img{B} > 0.2$, $l$ is always close to zero.
	The other case, when $\img{A}$ is white, is not as radical, but for nearly white images $\img{B}$,
	relatively large changes in $\img{B}$ do not change the SSIM value much.
	Section~\ref{sec_luminance} reveals several interesting,  
	nonintuitive results based on this diagram, and as a consequence,
	shows that the $l$ component is not perceptually based.

\section{Evaluation}
\label{sec_dissection}

All results showing MSSIM values and the images of SSIM index maps were
computed using the Matlab script (\url{www.cns.nyu.edu/~lcv/ssim/}) of Wang et al. All 
images are made available as supplemental material, as well as the scripts
to generate our results. Since images ideally should be viewed
on a display at 100\% scale, instead of in a PDF viewer or on paper,
we urge the reader to look at the images in our supplemental material.
$\mathrm{SSIM}$ is visualized using a heatmap where white is $\mathrm{SSIM}=1$ (identical images),
black is $\mathrm{SSIM}=0$, and SSIM values in $(-1,0]$ map to $(\mathrm{red},\mathrm{green}]$.

\subsection{Luminance}
\label{sec_luminance}
\begin{figure}[t]
	\centering
	\ssimA{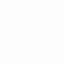}{253/255}{images/gray/gray255.png}{255/255}{0.99997}	
	\\
	\ssimA{images/gray/gray128.png}{128/255}{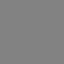}{130/255}{0.99988}	
	\\
	\ssimA{images/gray/gray000.png}{0}{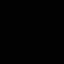}{2/255}{0.61914}	
	\\
	\ssimA{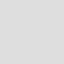}{222/255}{images/gray/gray255.png}{255/255}{0.99047}	
	\\
	\ssimA{images/gray/gray000.png}{0/255}{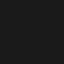}{26/255}{0.00953}	
	\caption{MSSIM behavior for constant grayscale images. The grayscale value is reported
		below each image. Note that in the third row,
		it is difficult to see any differences between the images,
		and still, MSSIM values indicate low quality. In the fourth row,
		it is clearly easy to see a difference between the images, while MSSIM indicate that
		they are similar.
		At the bottom, we see that when comparing a black image to an
		image with 26/255, MSSIM is almost zero.
	}
	\label{fig_luminance_weirdnesses}
\end{figure}

SSIM was designed so that $\mathrm{MSSIM}(\img{A},\img{A})=1$, that is, if the images
are the same, MSSIM will be one, and in general, a value $\geq 0.99$ indicates
that the images are indistinguishable.	
Here, we will explore how MSSIM behaves for images that only contain a single grayscale value.
For such images, $c=1$ and $s=1$, and therefore, it is only
the $l$ component that affects the values in this experiment.
In Figure~\ref{fig_luminance_weirdnesses} we reveal some results that have previously been
unknown (to the best of our knowledge).
The first row compares a white ($255/255$) image against a nearly-white ($253/255$)
image, and MSSIM is almost one, which makes sense, since it is hard to
see any difference between these two images. The difference in grayscale values
is $2/255$. On the second row, we do the same but for mid-gray images with difference
$2/255$ and the result is similar.
However, the third row compares a black image to a nearly-black ($2/255$)
image, again with a difference of 2/255, and in this case, MSSIM values are low, indicating
that the images are \textit{not} similar, when, in fact, it is	
difficult to see any difference between the two.

The fourth row is, perhaps, even more surprising, since MSSIM values indicate
that the images are similar, while they visually are not.
In the fifth row, the MSSIM values indicate that these two images are dissimilar,
and when increasing the value from 26 up to 255, the MSSIM value decreases to 0.0.
This means that for 90\% of the range, MSSIM is close to zero,
which unreasonably compresses the resolution of the index.
Mathematically, this stems from the quadratic forms in the normalization denominator
($\muA^2 + \muB^2 + C_1$) of Equation~\ref{eq_SSIM_luminance}, which exaggerates
differences near black.
Regardless, SSIM does not seem to be aligned with the HVS's ability
to detect luminance differences.
The orange and blue curves in Figure~\ref{fig_black_against_gray} predict
the results shown in Figure~\ref{fig_luminance_weirdnesses},
confirming that $l$ component of SSIM can be misleading.

\subsection{Color}
\label{sec_color_errors_ssim}
\begin{figure}[t]
	\centering
	\ssimA{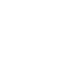}{white}{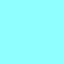}{\scriptsize{(0.56, 1, 1)}}{0.99047}
	\\
	\ssimA{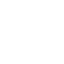}{white}{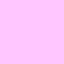}{\scriptsize{(1, 0.78, 1)}}{0.99047}
	\\
	\ssimA{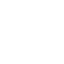}{white}{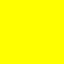}{\scriptsize{(1, 1, 0)}}{0.99276}
	\caption{Color images and SSIM do not work well together, as shown here.
		We have reduced the red (top), green (middle), and blue (bottom) color components until the error
		became $\approx\!0.99$.
		Numbers in parentheses represent RGB. Clearly, the differences are visible even though SSIM values indicate
		that the image pairs are extremely similar.
	}
	\label{fig_chrominance_weirdnesses}
\end{figure}

The SSIM authors present results using only JPEG and JPEG2000 color images~\cite{Wang2004a},
but add that using other color components does not significantly change the performance of the model.	
The index has still, nevertheless, been used on color images
by first converting to grayscale values, using, for example, the color encoding standard Rec.~601 (which is used in
Matlab's \texttt{rgb2gray}, and is the procedure recommended on the SSIM website):
\begin{equation}
Y = 0.2989r + 0.5810g + 0.1140b.
\label{eq_luminance_conversion}
\end{equation}
Another approach is to convert from $RGB$ to $YC_rC_b$, and apply SSIM to
$Y$, $C_r$, and $C_b$ individually, and then use 
$0.8\mathrm{SSIM}_Y + 0.1\mathrm{SSIM}_{C_r} + 0.1\mathrm{SSIM}_{C_b}$~\cite{Wang2004b}
as the quality index.

It is well-known that there are many colors of equal luminance and furthermore that
any mapping between $RGB$ and grayscale value is many-to-one.
As a consequence, SSIM can generate a high value, indicating similarity, even though the colors are visibly dissimilar.
This is visualized for three color pairs in Figure~\ref{fig_chrominance_weirdnesses}, where the original \texttt{rgb2gray}
function in Matlab has been used.
It is evident that simply converting from RGB to grayscale values can give erroneous
SSIM results, as would any metric relying on a reduction function from color to grayscale.
Even the original SSIM paper~\cite{Wang2004a} does this, and based on our findings here,
we advise not to use SSIM with color images.
A better solution would be to use a metric that inherently
handles color.

%
%
%
%

\begin{figure}[tb]
	\centering
	\ssimB{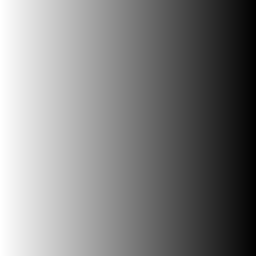}{\img{A}}{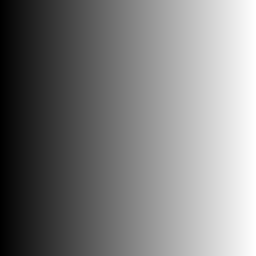}{\img{B}}{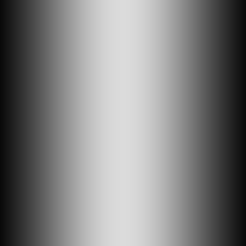}{0.51} 
	\hspace{3mm}\footnotesize{$256\times 256$}  \\
	\ssimB{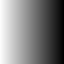}{\img{A}}{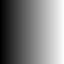}{\img{B}}{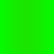}{-0.07}
	\hspace{6mm}\footnotesize{$64\times 64$}  \\
	\ssimB{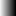}{\img{A}}{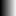}{\img{B}}{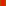}{-0.82}
	\hspace{6mm}\footnotesize{$16\times 16$}
	\vspace{-2mm}
	\caption{SSIM images of gradients versus mirrored gradients at different resolutions, shown to the right.
		Grayscale indicates positive SSIM, while green and red indicate negative (the same heatmap is used here
		as in Figure~\ref{fig_teaser}). The numbers under the
		images are the MSSIM values.
		Note that the original SSIM implementation, uses only ``valid'' values for
		the filter, which means that it removes a border of 5 pixels around the SSIM image.
		Hence, the bottom SSIM image is only $16-5-5=6$ pixels wide, for example.
	}
	\label{fig_gradients}
\end{figure}

\subsection{Gradients}
Next, we compare
an image containing a gradient against a horizontally mirrored
version of the same image at different resolutions.
The results are summarized in Figure~\ref{fig_gradients}.
In all examples, $c=1$ and the image of $l$ (not shown) contains vertical lines of constant values, starting
with a low value at the left, peaking in the middle, and going down to a low value at the right.
The $s$ component, which is not shown in the figure, starts at $0.86$ for the pair with $256\times 256$ pixels, but goes down
to $-0.10$ for the middle row ($64\times 64$), and even further down to $-0.90$
for $16\times 16$ pixels, which means that it is the $s$ component that makes MSSIM values in 
Figure~\ref{fig_gradients} negative toward the bottom.
This was surprising since the $\img{A}$ images
are similar at all resolutions, as are the $\img{B}$ images.
Assuming the $256\times 256$ and the $16\times 16$ gradients are part of a large image,
the perceived error in the $16\times 16$ region will surely be less glaring than in the $256\times 256$ region.
This is the opposite of the results, generated by SSIM, shown in Figure~\ref{fig_gradients}.

Recall that a negative $s$ to the power of a non-integer value generates a complex number
or an undefined result (see the last part of Section~\ref{sec_min_values}), and this problem
will occur for the two bottom rows in Figure~\ref{fig_gradients}.
These examples might seem overly contrived, but serve to demonstrate that it is indeed possible, and not highly
unlikely, for SSIM to generate negative values for simple distortions (see also Figure~\ref{fig_teaser}\textit{c},
~\ref{fig_teaser}\textit{e}, and~\ref{fig_teaser}\textit{f},
which are discussed below).

\begin{figure*}
	\centering
	\setlength{\fboxsep}{0pt}%
	\setlength{\fboxrule}{0.2mm}%
	\setlength{\tabcolsep}{0pt}
	\renewcommand{\arraystretch}{0.2}
	\newcommand{\teaserImageHeight}{0.13\textwidth}
	\newcommand{\smallTeaserImageHeight}{0.0418\textwidth}
	\footnotesize
	\begin{tabular}{cccccccccccccccccccccccc}
		& \multicolumn{3}{c}{\textit{a}} & \multicolumn{3}{c}{\textit{b}} & \multicolumn{3}{c}{\textit{c}}
		& \multicolumn{3}{c}{\textit{d}} & \multicolumn{3}{c}{\textit{e}} & \multicolumn{3}{c}{\textit{f}} & \multicolumn{3}{c}{\textit{g}} &
		\\
		\rotatebox{90}{\hspace*{3.2mm}reference}
		& \multicolumn{3}{c}{\includegraphics[height=\teaserImageHeight]{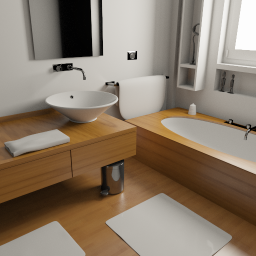}} 
		& \multicolumn{3}{c}{\includegraphics[height=\teaserImageHeight]{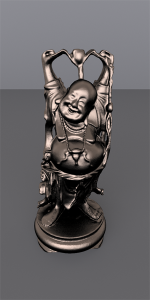}}
		& \multicolumn{3}{c}{\includegraphics[height=\teaserImageHeight]{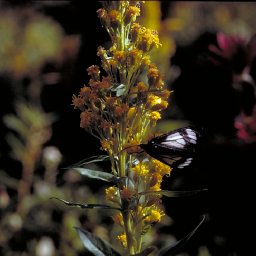}}
		& \multicolumn{3}{c}{\includegraphics[height=\teaserImageHeight]{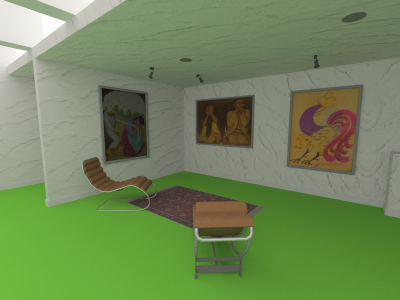}}
		& \multicolumn{3}{c}{\includegraphics[height=\teaserImageHeight]{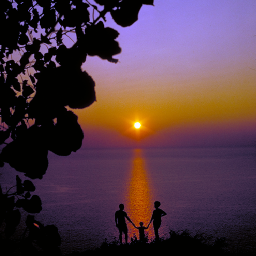}}
		& \multicolumn{3}{c}{\includegraphics[height=\teaserImageHeight]{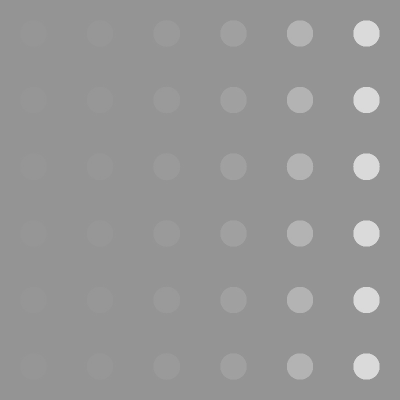}}
		& \multicolumn{3}{c}{\includegraphics[height=\teaserImageHeight]{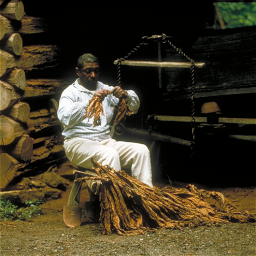}}
		&
		\\ 
		\rotatebox{90}{\hspace*{8mm}test}
		& \multicolumn{3}{c}{\includegraphics[height=\teaserImageHeight]{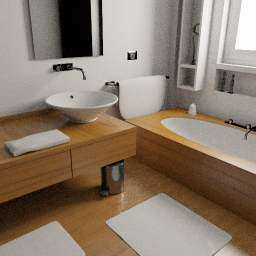}}
		& \multicolumn{3}{c}{\includegraphics[height=\teaserImageHeight]{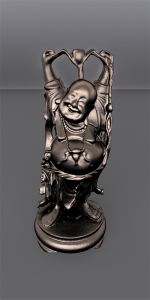}}
		& \multicolumn{3}{c}{\includegraphics[height=\teaserImageHeight]{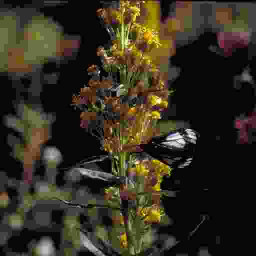}}
		& \multicolumn{3}{c}{\includegraphics[height=\teaserImageHeight]{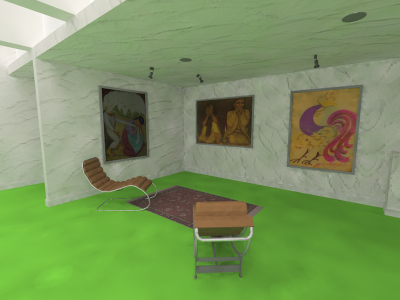}}
		& \multicolumn{3}{c}{\includegraphics[height=\teaserImageHeight]{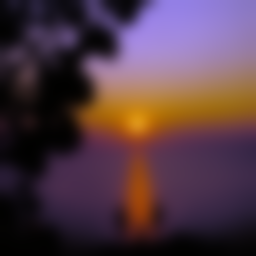}}
		& \multicolumn{3}{c}{\includegraphics[height=\teaserImageHeight]{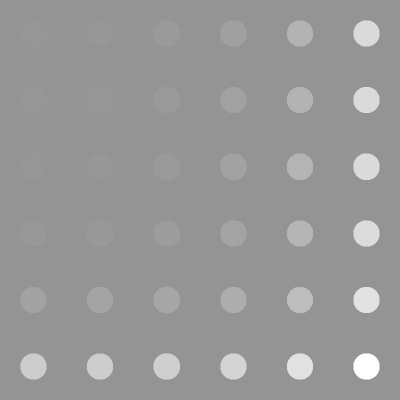}}
		& \multicolumn{3}{c}{\includegraphics[height=\teaserImageHeight]{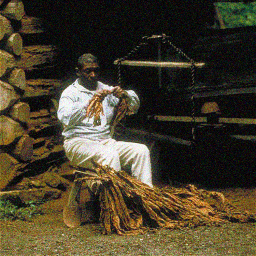}}
		&
		\\ 
		\rotatebox{90}{\hspace*{5mm}SSIM}
		& \multicolumn{3}{c}{\includegraphics[height=\teaserImageHeight]{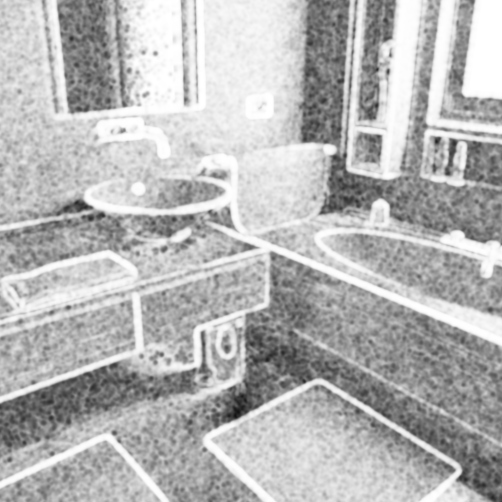}}
		& \multicolumn{3}{c}{\includegraphics[height=\teaserImageHeight]{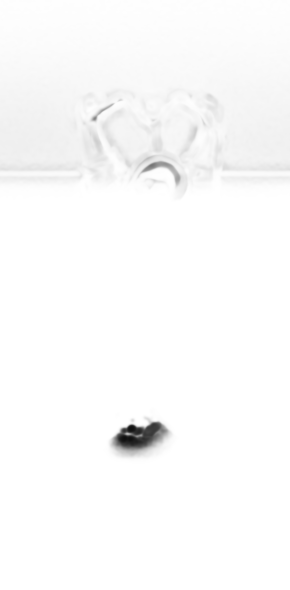}}
		& \multicolumn{3}{c}{\includegraphics[height=\teaserImageHeight]{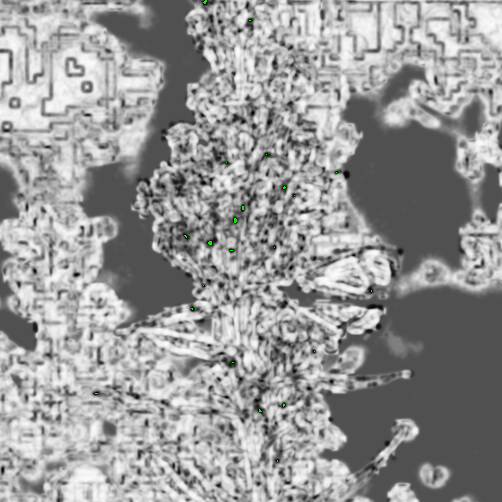}}
		& \multicolumn{3}{c}{\includegraphics[height=\teaserImageHeight]{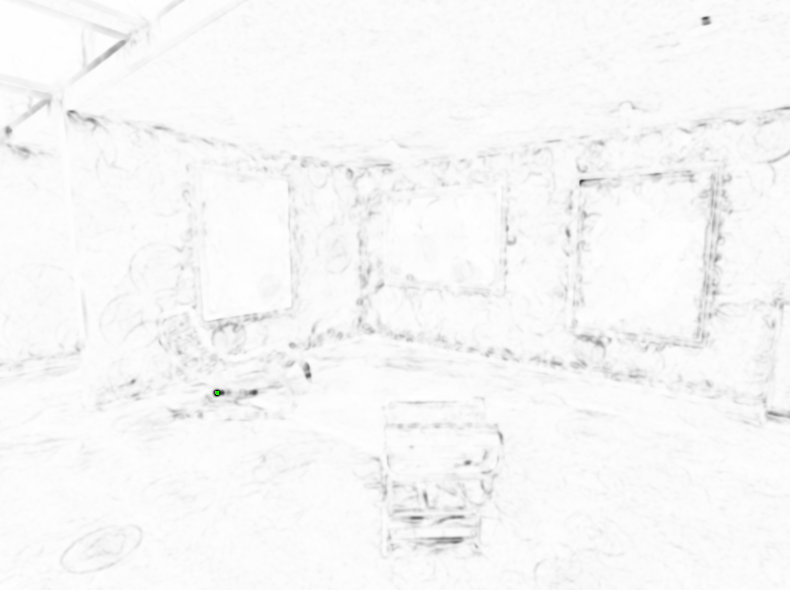}}
		& \multicolumn{3}{c}{\includegraphics[height=\teaserImageHeight]{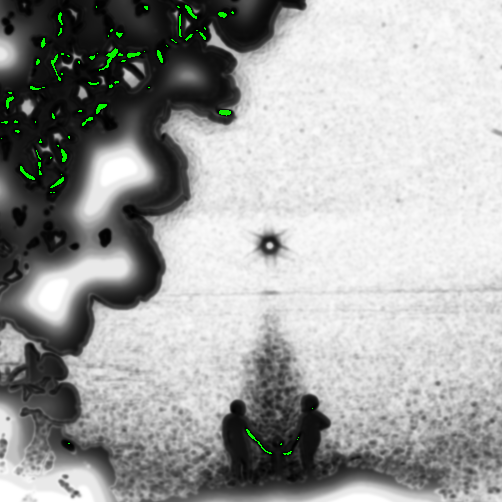}}
		& \multicolumn{3}{c}{\includegraphics[height=\teaserImageHeight]{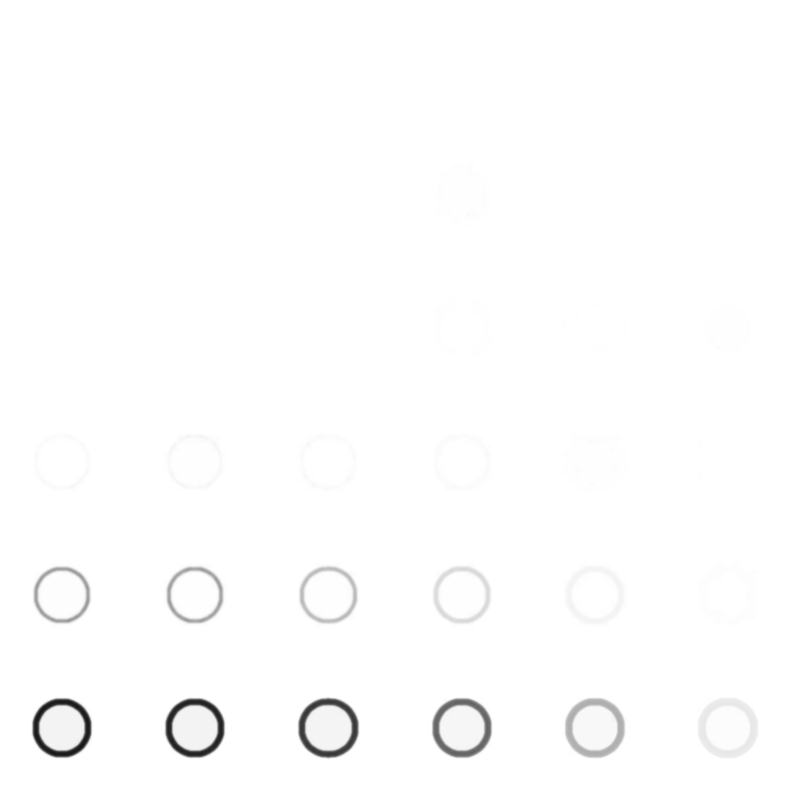}}
		& \multicolumn{3}{c}{\includegraphics[height=\teaserImageHeight]{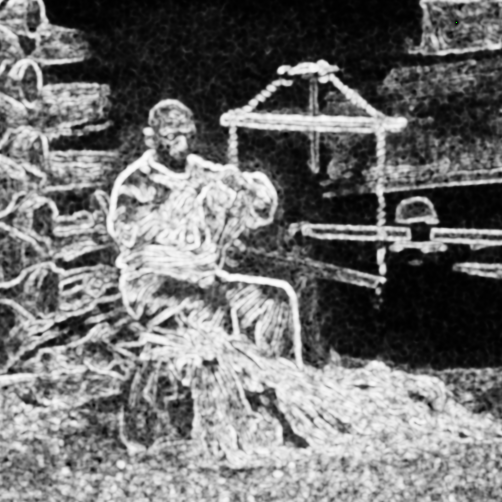}}
		& \fbox{\includegraphics[height=\teaserImageHeight]{images/teaser/heatmap.png}}
		\\ 
		& \fbox{\includegraphics[height=\smallTeaserImageHeight]{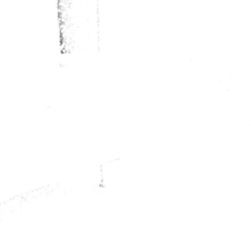}}
		& \fbox{\includegraphics[height=\smallTeaserImageHeight]{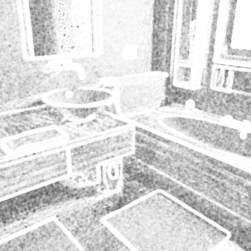}}
		& \fbox{\includegraphics[height=\smallTeaserImageHeight]{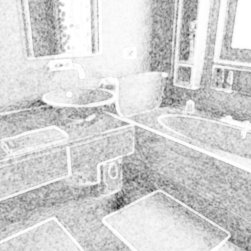}}
		& \fbox{\includegraphics[height=\smallTeaserImageHeight]{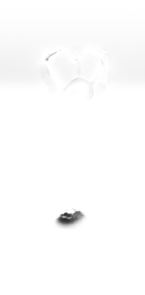}}
		& \fbox{\includegraphics[height=\smallTeaserImageHeight]{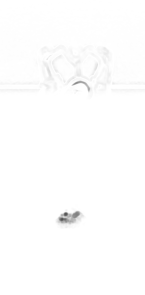}}
		& \fbox{\includegraphics[height=\smallTeaserImageHeight]{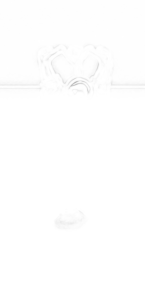}}
		& \fbox{\includegraphics[height=\smallTeaserImageHeight]{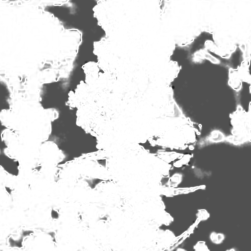}}
		& \fbox{\includegraphics[height=\smallTeaserImageHeight]{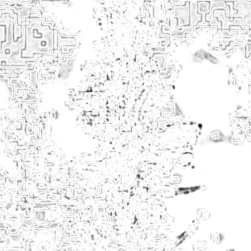}}
		& \fbox{\includegraphics[height=\smallTeaserImageHeight]{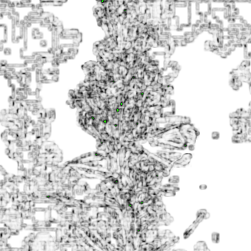}}
		& \fbox{\includegraphics[height=\smallTeaserImageHeight]{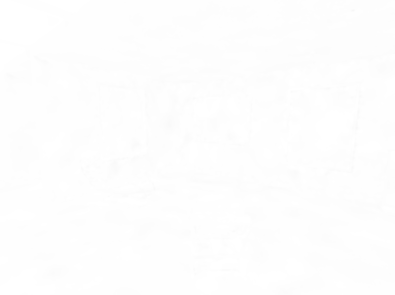}}
		& \fbox{\includegraphics[height=\smallTeaserImageHeight]{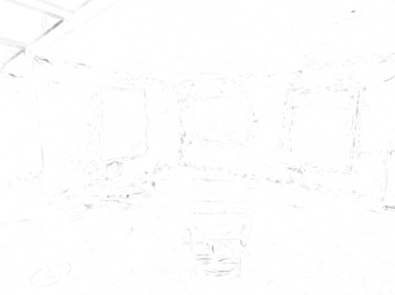}}
		& \fbox{\includegraphics[height=\smallTeaserImageHeight]{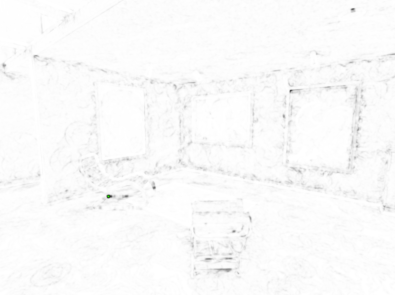}}
		& \fbox{\includegraphics[height=\smallTeaserImageHeight]{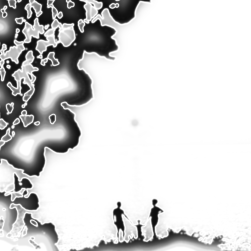}}
		& \fbox{\includegraphics[height=\smallTeaserImageHeight]{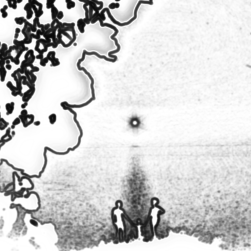}}
		& \fbox{\includegraphics[height=\smallTeaserImageHeight]{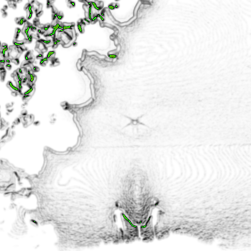}}
		& \fbox{\includegraphics[height=\smallTeaserImageHeight]{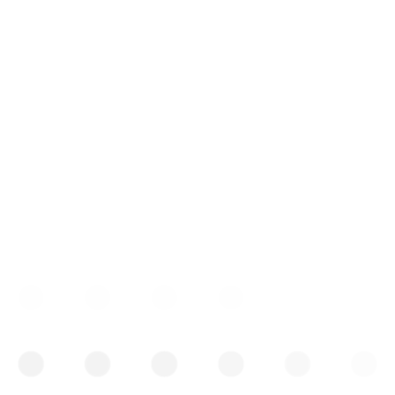}}
		& \fbox{\includegraphics[height=\smallTeaserImageHeight]{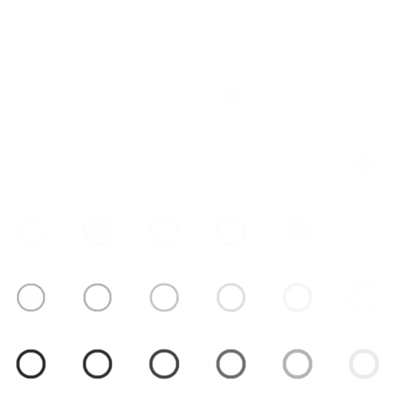}}
		& \fbox{\includegraphics[height=\smallTeaserImageHeight]{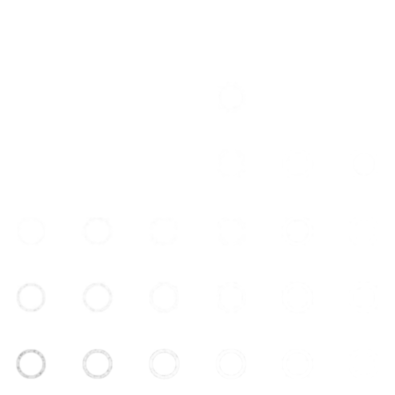}}
		& \fbox{\includegraphics[height=\smallTeaserImageHeight]{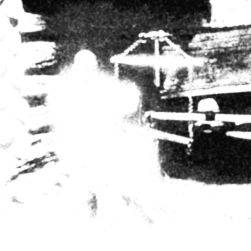}}
		& \fbox{\includegraphics[height=\smallTeaserImageHeight]{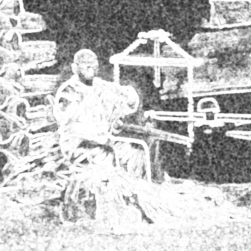}}
		& \fbox{\includegraphics[height=\smallTeaserImageHeight]{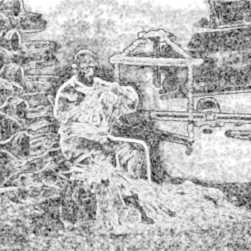}}
		&
		\\
		&
		$l$ & $c$ & $s$ &
		$l$ & $c$ & $s$ &
		$l$ & $c$ & $s$ &
		$l$ & $c$ & $s$ &
		$l$ & $c$ & $s$ &
		$l$ & $c$ & $s$ &
		$l$ & $c$ & $s$ &
	\end{tabular}
	\caption{
		Images should ideally be viewed
		on a display at 100\% scale, so
		we urge the reader to look at the images in our supplemental material.
		The same heatmap is used here as in Figure~\ref{fig_teaser}. All image pairs are discussed in Section~\ref{sec_complex_images}.
		Note that all reference and test images, but not SSIM images, are reduced in size for space considerations.
	}
	\label{fig_complex_images}
\end{figure*}

\begin{figure}[tb]
	\setlength{\tabcolsep}{1pt}
	\renewcommand{\arraystretch}{0.5}
	\begin{tabular}{ccc}
		\includegraphics[width=0.32\columnwidth]{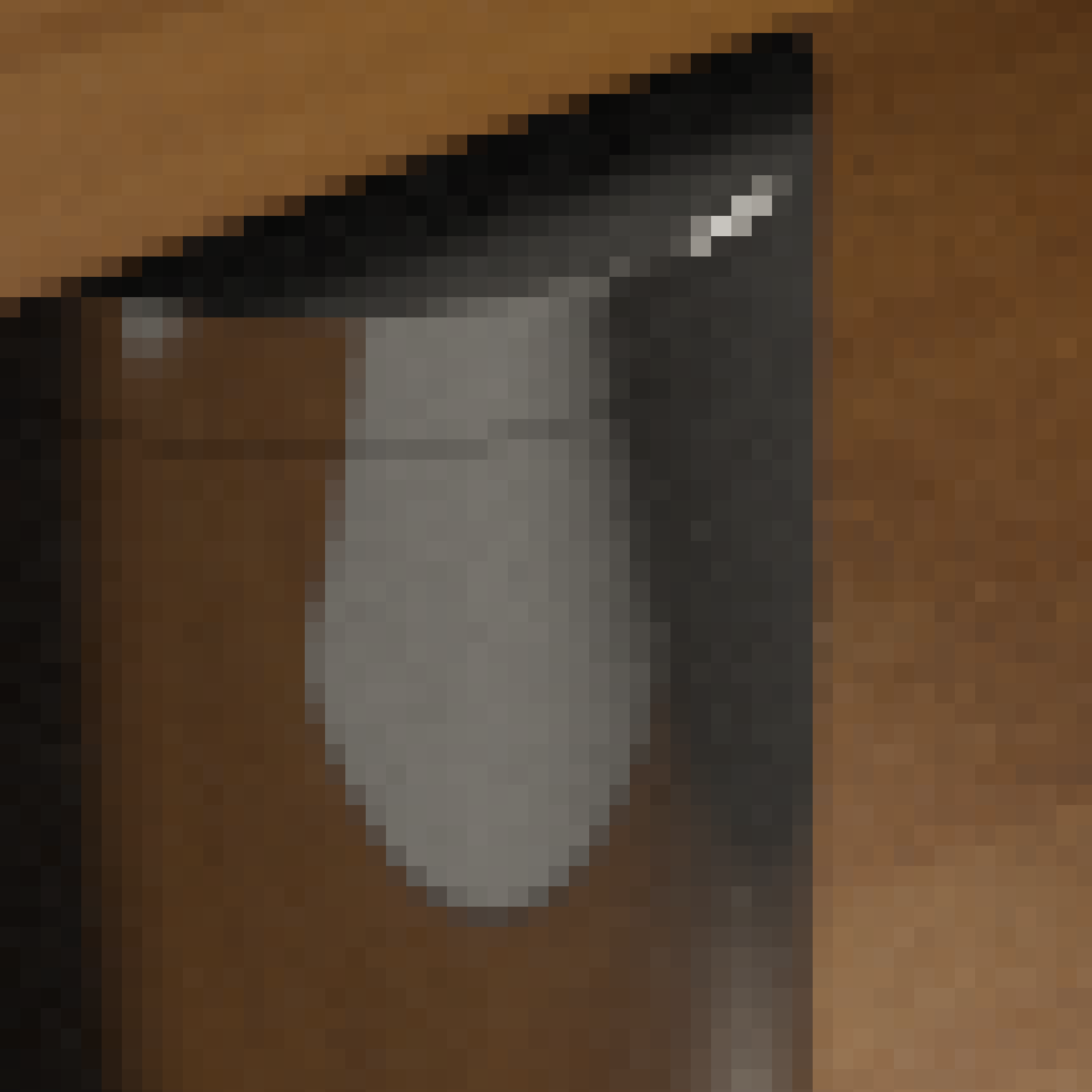} &
		\includegraphics[width=0.32\columnwidth]{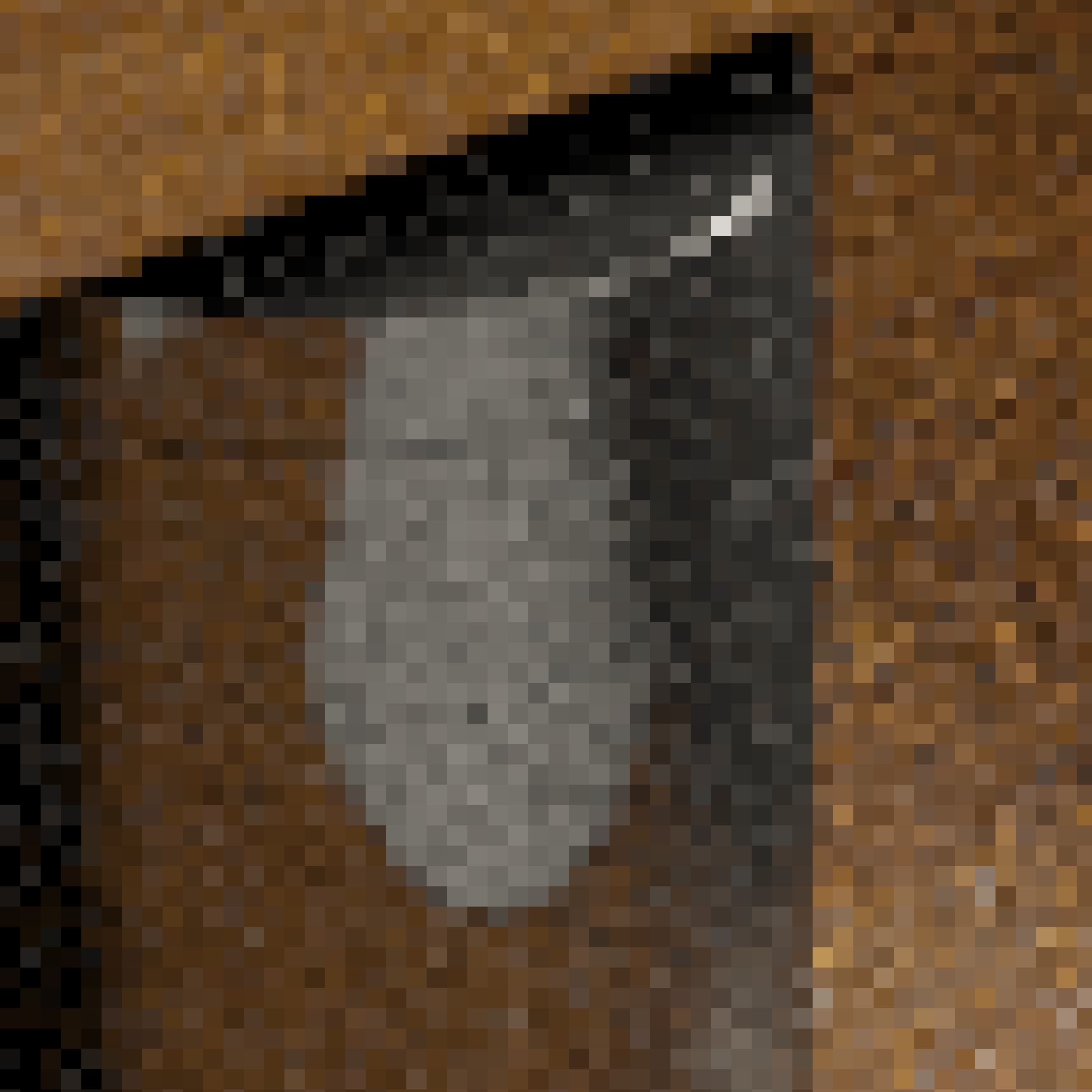} &
		\includegraphics[width=0.32\columnwidth]{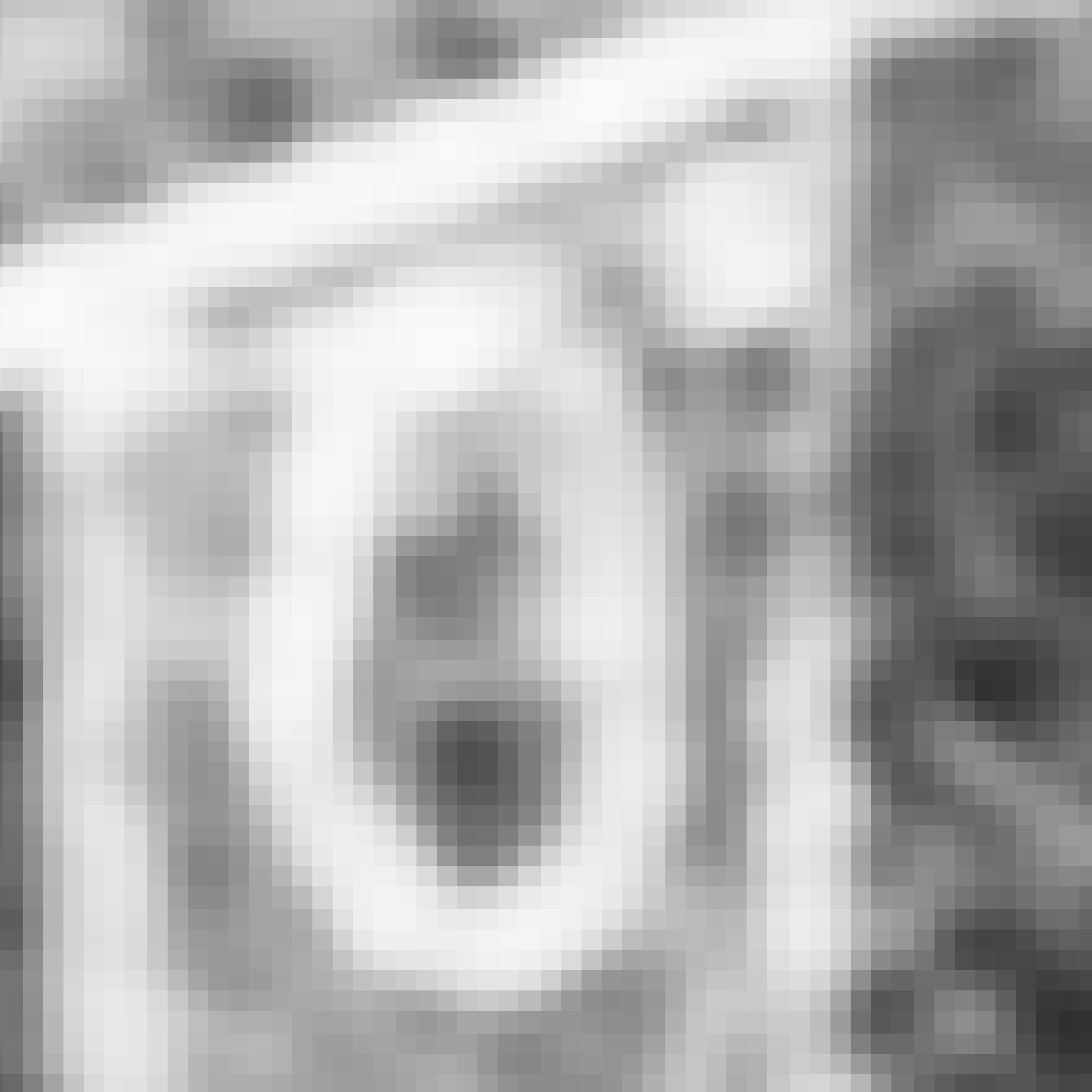}
	\end{tabular} 
	\caption{Zooming in on the trash can in Figure~\ref{fig_complex_images}\textit{a}, we see that SSIM values
		tends to have very high values, 
		i.e., low error, in a relatively large vicinity of contrast edges, which is unexpected since
		the Monte Carlo noise is of the same magnitude on both sides.
	}
	\label{fig_zoomed_edges}
\end{figure}

\subsection{Complex Images}
\label{sec_complex_images}

The average MSSIM value has been correlated with the subjective quality (five levels, from ``Bad'' to ``Excellent'')
of JPEG-compressed images~\cite{Wang2004a}. The resulting SSIM image has however, to our knowledge, never been
evaluated with subjective observers.
It is certainly true that SSIM in many cases finds image differences that are also detected by a human
observer. Had it not, its use would have been less widespread.
With well-behaved image pairs, 
the corresponding SSIM map tells a convincing story.
Synthetic examples (seen above, for example) serve to clarify situations where SSIM behaves contradictory to human perception,
and it can be challenged whether these situations arise for more general images with reasonable distortions.
In the following, we will demonstrate several such cases.
All observations have been carried out on a Dell UP3216Q monitor, calibrated
for sRGB (IEC 61966-2-1:1999 standard), with a D65 standard illuminant. 

We start by discussing the images in Figure~\ref{fig_teaser} in more detail.
In image pair \textit{a}, the surface of the moon has substantially different luminance,
but that error is detected as lower than the text ``Houston, we have a problem!', which is
difficult to see for most people. 
As we have seen earlier, this can be explained using
Figure~\ref{fig_black_against_gray}. 
The test image in \textit{b} has its chrominance
shifted compared the reference, and it is clear that
SSIM does not react much to this, and the explanation to this
is given in Section~\ref{sec_color_errors_ssim}. 
Image pair \textit{c} is from the LocVis~\cite{Wolski2018} database and shows
a situation where the SSIM values are high on the red door and on the red
window shutters, while it is clear that there are differences.
The explanation is that after grayscale conversion, the grayscale values
on the door and window shutters are the same in the reference and test images.
%

Image pair \textit{d}, from the CS-IQ database~\cite{Larson2010}, shows that SSIM
sometimes can generate high SSIM values on edges in images, even though the
error is evenly spread out over the entire image. As can be seen below the
SSIM images, the $l$, $c$, and $s$ components are all white on edges, so none of the terms detect the induced error.
In image pair \textit{e}, we have introduced a dithering that in a checkerboard pattern
adds and subtracts 6 from the pixel (8-bit) grayscale value, clamped to 0 and 255, respectively.
The test image uses the same, but inverse, dithering.
To the human observer, we claim that these distortions have little, or no, visual impact.
SSIM, however, finds these two images very dissimilar with many pixels generating negative 
(green) SSIM values. As seen in Section~\ref{sec_min_values}, a negative value
to the power of a floating-point exponent 
is undefined in most programming languages or otherwise generates a complex number.
Again, this is difficult to interpret in an image quality measure.
The images in Figure~\ref{fig_teaser}\textit{f}, with lowered contrast compared to reference,
also from the CS-IQ database, show an SSIM image that
is particularly counterintuitive---in the bright regions, where visible differences can
be argued to be largest, the SSIM image is bright as well, while in the dark regions of the reference and test
image, SSIM values are much lower. As can be seen below the SSIM image, it is mostly the $l$
component that makes this so, which has been explained in Section~\ref{sec_luminance}.

Turning to Figure~\ref{fig_complex_images}, image pair \textit{a} is a path traced rendering with different sample
counts per pixel, for which the SSIM values at first seems to correspond well with the actual image differences.
Closer inspection reveals two important exceptions, namely, high similarity along all high contrast edges, and an
unreasonable dissimilarity for darker regions, mainly attributed to the $c$ component.
In Figure~\ref{fig_zoomed_edges}, we have zoomed in on the trash can in these images, to further illustrate
how large the region around an edge is, with very high SSIM values. 
Image pair \textit{b}, from LocVis, with different lighting conditions for reference and test, is noteworthy, since
it demonstrates both a false positive (spot under the Buddha, nearly invisible) and a false negative
(failure to detect luminance shift in the upper background) in the same image.
The JPEG-compressed test image in \textit{c}, from CS-IQ, is highly distorted in the detailed regions.
A slight difference in gray level in the dark regions between the reference and test images, however,
dominates the final output. Additionally, a few negative values show up for the $s$ component.
%

Another example of what we consider is a false negative, is the blotchy distortion of image \textit{d} (from LocVis),
which most notable on the green floor.  SSIM fails to detect this disturbing variation in intensity.
Turning to the heavily blurred test image in \textit{e}, from CS-IQ, the different components of SSIM again
highlights dark regions where the images only differ by a small amount. 
The large drop in intensity of the sun is nearly ignored and the indication of high similarity
in its middle is exaggerated.
For the synthetic image pair in \textit{f}, from LocVis, the $c$ component shows dissimilarity around the
edges of the circles, while the $l$ component, containing information about the intensity
levels within the circles, shows high similarity. We consider the lower left circle in the test image to be
significantly different from the reference image, but it does not show up much inside the circle.
The final image pair \textit{g}, from CS-IQ, is reported as strikingly dissimilar by SSIM across all dark regions, 
which is indicated by both the
$l$ and $c$ components.  The additive noise distortion is spread evenly across the
whole test image, but in our opinion, these errors are harder to detect in darker areas, which 
seems to be the inverse behavior of SSIM\@.




%

\section{Conclusions}
\label{sec_conclusions}

We have demonstrated the mathematical properties of SSIM and shown that it is
not adhering to properties of the human visual system. 
This is not surprising, as it was not a goal stated in the original work~\cite{Wang2002} 
upon which SSIM is based.	
However, over time, the similarity index has grown in number of uses, and
popular belief of the index's capabilities has significantly widened with respect
to this original scope. 
The purpose of this paper is to moderate this belief, since it can guide research in the wrong direction.
Even though SSIM generates useful results in some cases, it can generate counterintuitive
results in many others, as we have seen.
This opens the door to research improved metrics to describe how humans detect differences
between images, be they synthetic, rendered, or natural.

SSIM is at its core a statistical measure~\cite{Wang2002},
a product of three local dissimilarity factors, namely, luminance, variance, and correlation.
We have derived these factors' minima and shown how
their ranges and normalization are creating nonintuitive results. This occurs, for example, 
for low luminance values or when the local distribution of pixel values visually
differ very little, though regularly.  We have also shown that the original SSIM formulation with certain parameters	
can output undefined results.
For one of the major areas of use for SSIM,
namely rendering, these results constitute the core of the index's
weaknesses---both as a qualitative indication, using the pooled MSSIM value,	
and as a quantitative value, when using the SSIM map to understand
the visual performance of rendering algorithms.

Current graphics and rendering research has a major focus on Monte Carlo ray tracing
and denoising and reconstruction algorithms using neural networks.
Such networks often introduce small variations during training and could  
potentially suffer disproportionately
from the shortcomings of SSIM\@. We thus encourage further graphics research to employ
the index with care and caution, or preferably replace it, since its use may distort
or bias image quality assessment.
The \textit{difference evaluator for alternating images} (\FLIP)~\cite{Andersson2020}, is a step toward such a replacement.

	\section*{Acknowledgements}
	We would like to thank Magnus Andersson, Robert Toth, Eric Haines, Matt Pharr, Eric Endterton, and Aaron Lefohn, for their valuable comments and feedback on this work.
	
	\newpage
	
	\bibliographystyle{acm}
	{\small\bibliography{references}}
\end{document}